\documentclass[conference,10pt,letterpaper]{IEEEtran}
\usepackage{amsmath,amssymb,latexsym}
\usepackage{graphicx}
\usepackage{subfigure}
\usepackage[noadjust]{cite}
\usepackage{epsfig}
\usepackage{verbatim}
\usepackage{arcs}
\usepackage{setspace}
\usepackage{bm}
\usepackage{yhmath}
\usepackage[ruled,lined,linesnumbered]{algorithm2e}
\usepackage{algpseudocode}

\IEEEoverridecommandlockouts
\newcommand{\ca}{\mathcal}
\newtheorem{df}{Definition}

\newtheorem{ps}{Proposition}
\newtheorem{lm}{Lemma}

\title{Delay-Optimal Distributed Edge Computing in Wireless Edge Networks}

\author{\IEEEauthorblockN{Xiaowen Gong \\ }
\IEEEauthorblockA{Department of Electrical and Computer Engineering\\
Auburn University, Auburn, AL 36849\\
Email: xgong@auburn.edu\\}
\thanks{The work of Xiaowen Gong is supported by his startup fund and Intramural Grants Program 190599 provided by Auburn University.}
}

\begin{document}

\maketitle
\thispagestyle{empty}
\pagestyle{empty}

\begin{abstract}


By integrating edge computing with parallel computing, distributed edge computing (DEC) makes use of distributed devices in edge networks to perform computing in parallel, which can substantially reduce service delays. In this paper, we explore DEC that exploits distributed edge devices connected by a wireless network to perform a computation task offloaded from an end device. In particular, we study the fundamental problem of minimizing the delay of executing a distributed algorithm of the computation task. We first establish some structural properties of the optimal communication scheduling policy. Then, given these properties, we characterize the optimal computation allocation policy, which can be found by an efficient algorithm. Next, based on the optimal computation allocation, we characterize the optimal scheduling order of communications for some special cases, and develop an efficient algorithm with a finite approximation ratio to find it for the general case. Last, based on the optimal computation allocation and communication scheduling, we further show that the optimal selection of devices can be found efficiently for some special cases. Our results provide some useful insights for the optimal computation-communication co-design. We evaluate the performance of the theoretical findings using simulations.




\end{abstract}

\section{Introduction}\label{sc:intr}


Edge computing has recently emerged as a promising paradigm that performs a substantial amount of computing, storage, networking, and management functions on devices at or close to end users (referred to as ``edge devices''). This trend is largely enabled by the proliferation of smart devices with powerful computing capabilities, which are often not fully utilized. Compared to cloud computing which performs computing in remote data centers, edge computing can offload a large amount of data traffic from the core network to the edge network, which can greatly reduce the \textit{communication delay} incurred in the network. One main driving force for the popularity of edge computing is many emerging applications of AI that require very low service delays, including augmented reality (AR)~\cite{googleglass}, virtual reality (VR)~\cite{oculus}, and autonomous vehicle~\cite{uav}. These applications are empowered by recent advances in machine learning (ML), which typically rely on computationally intensive processing of large amounts of data.


On the other hand, distributed computing is a traditional computing paradigm that uses distributed devices to perform computing \textit{cooperatively in parallel}. One main advantage of this approach is that it can greatly reduce \textit{computation delay} by parallelizing the algorithm of computation and distributing the computation workload to different devices, rather than performing all the computation on a single device. One well-known
example of distributed computing is cloud computing in data centers, which utilizes large clusters of interconnected computer servers for parallel computing.



To exploit the potential of edge computing, it is promising to leverage \textit{distributed edge computing} (DEC), which makes use of distributed \textit{edge devices} to perform computing cooperatively in parallel. This approach is enabled by widely available edge devices with under-utilized computing capacities that can be connected by wireless networks in a distributed manner. To accelerate many emerging applications that require very low delays, it is beneficial to \textit{offload and distribute} a large computation workload from a single end device to possibly multiple edge devices nearby. One main advantage of this approach is to perform the computation in parallel which reduces the computation delay. If a computation is offloaded from one device and distributed to $N$ devices with the same computing power, and if communication delays are not counted (which certainly should and will be considered), then the computation delay will reduce by a fold of $N$, which is very appealing. Another advantage of that approach is to leverage an edge device(s) with higher computing power to reduce the computation delay. This will happen more likely when dedicated and powerful edge computer servers are deployed as infrastructure in the future.

In this paper, we study DEC that exploits wirelessly connected edge devices to perform distributed computing. Our goal is to minimize the delay of executing a distributed algorithm. To this end, we will study the optimal allocation of computation workloads of the distributed algorithm (referred to as ``computation allocation'') to devices. We will also investigate the optimal scheduling of communications between the devices in the wireless network. We will further study the optimal selection of devices for executing the distributed algorithm. We will explore fundamental issues in the cross-layer design, analysis, and optimization of DEC in wireless networks.


The computation allocation and communication scheduling for DEC are significantly different from prior studies. These two problems have non-trivial coupling, as the optimal solution to each problem depends on the solution to the other problem. Therefore, compared to existing works on distributed computing, the computation allocation here needs to take into account the features of wireless networks, including \textit{interference} among wireless links and \textit{diverse data rates} of wireless links. Moreover, the objective of communication scheduling here is to minimize the delay of executing a distributed algorithm, which is quite different from existing studies on wireless network scheduling.

The main contributions of this paper can be summarized as follows.
\begin{itemize}
  \item We propose a framework of distributed computing with a parallel algorithm structure using edge devices connected by a wireless network. Based on this framework, we formulate the problem of allocating computation workloads to devices and scheduling communications between devices for minimizing the delay of executing the distributed algorithm.

  \item We first establish some structural properties of the optimal communication scheduling, which show that it is optimal to be non-preemptive, be non-idle, and schedule forward communications before backward communications. Then, given these properties, we characterize the optimal computation allocation by developing an efficient algorithm to find it. Next, based on the optimal computation allocation, we characterize the optimal scheduling order of communications for the cases with uniform communication delays or computation rates. We also develop an efficient algorithm with a finite approximation ratio that finds the optimal scheduling order for the general case with diverse communication and computation rates. Based on the optimal computation allocation and communication scheduling, we further show that the optimal selection of devices can be found by an efficient linear search for the cases with uniform communication delays or computation rates. Our results provide some useful insights for the optimal computation-communication co-design.
 
  
  \item We evaluate the performance of the optimal polices using simulations. The simulation results demonstrate that the optimal polices outperform non-optimal policies, and are more advantageous when communication delays and/or computation rates are more diverse.
  

\end{itemize}

The rest of this paper is organized as follows. Section \ref{sc:rel} reviews related work. In Section \ref{sc:mod}, we describe a framework of distributed edge computing in wireless networks. Section~\ref{sc:alloc_sched} focuses on the optimal computation allocation and communication scheduling. Simulation results are discussed in Section \ref{sc:sim}. Section \ref{sc:con} concludes this paper and discusses future work.


\section{Related Work}\label{sc:rel}


\noindent\textbf{Edge computing.} Edge computing has attracted growing research interests in the past few years~\cite{chiang16}. Many works have used edge devices for video applications that require low service delays, such as for video rendering~\cite{wu17}, and virtual reality~\cite{zhang17,wang18}. One important application studied in these works is real-time video inference~\cite{liu18,ran18,cheng18}. Computation offloading from mobile devices to edge servers has also been studied~\cite{chen15,tong16,tan17,gao19}. Another major research direction of edge computing is edge caching~\cite{yang17}. Learning users' interested contents for caching has also been studied~\cite{wang14}. Cooperative networks of caches have also been studied~\cite{ao15,liu15}. However, existing works on edge computing have not considered offloading and distributing computation to more than one device, with the goal of reducing the computation delay. 




\noindent\textbf{Distributed computing.} There have been many studies on the design of distributed algorithms and computation allocation for reducing computation delays~\cite{ross91,chang11,zheng13,zhu14,zheng15,im16}. Some of these works have studied the effects of the network on computing~\cite{cheng90,tan13,jiang16}. Some other works have considered the throughput of networked computers for processing computations~\cite{xie16}. Recent studies have considered the cross-layer design of distributed computing and networking for improving computation delay~\cite{chen12} or throughput~\cite{liu13,ghaderi15}. On the other hand, many works have studied communication scheduling in data center networks~\cite{shafiee17}. A large body of these works have focused on the scheduling of co-flows under distributed computing frameworks, in particular MapReduce~\cite{chowdhury11,zhao15,li16}. However, most existing works on distributed computing have not considered offloading and distributing computation to more than one device, with the goal of reducing the computation delay. Moreover, many studies have considered wired networks of devices, which do not take into account the features of wireless networks, including interference among wireless links.


\noindent\textbf{Wireless network scheduling.} Wireless network scheduling has been studied extensively for more than a decade. Most of the works have focused on the throughput of wireless networks~\cite{eryilmaz05}, including recent works on deadline-constrained throughput~\cite{hou09} and with distributed scheduling~\cite{jiang10}. Many other works have considered the total utility of data flows in the network~\cite{liu03} which depends on the throughput. Much fewer works have studied the delay performance of wireless network scheduling~\cite{gupta11}. On the other hand, some works have studied the cross-layer design of scheduling, routing, and/or congestion control for various objectives, including for throughput~\cite{tassiulas92,lin09}, delay~\cite{xue13}, or utility~\cite{yi08}. However, existing works on wireless network scheduling have not considered using distributed devices connected wirelessly to perform computation cooperatively, with the goal of reducing the computation delay.


\section{System Model and Problem Formulation}\label{sc:mod}




In this section, we first describe a framework of distributed computing with a parallel algorithm structure using wirelessly connected edge devices. Based on this framework, we then formulate the problem of minimizing the delay of executing the distributed algorithm. 




\noindent\textbf{Distributed algorithm.} Our goal is to execute an algorithm which consists of some computations. The algorithm can be executed by a single device in a centralized manner by performing the computations of the algorithm sequentially in some order. Alternatively, the algorithm can be executed in a distributed manner such that the computations are performed by distributed devices in parallel. By exploiting the computing power of distributed devices, this parallelization can greatly reduce the delay of executing the algorithm. 

\begin{figure}
\centering
\includegraphics[width=0.2\textwidth]{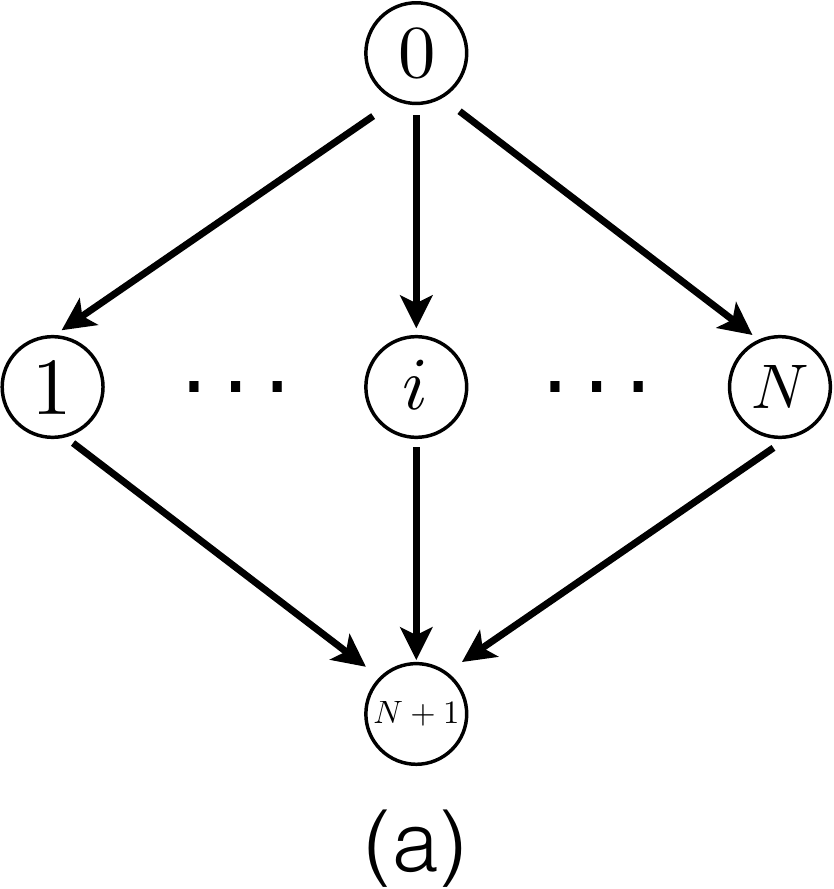}
\includegraphics[width=0.2\textwidth]{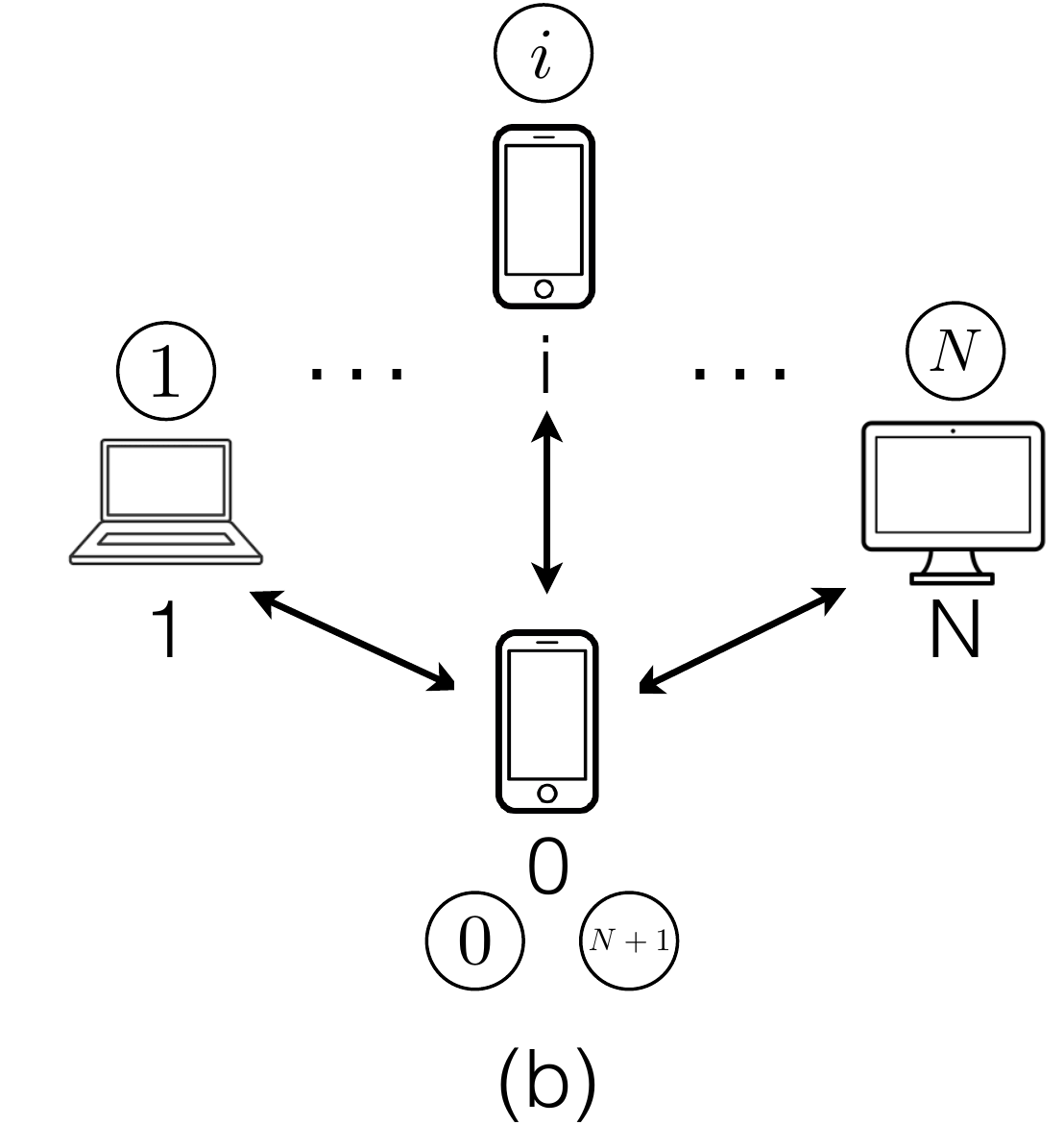}
\caption{(a) A  distributed algorithm with a parallel structure. Each node represents a computation; each directed edge represents a communication. (b) The computations of the distributed algorithm are allocated to devices connected by wireless links.}
\label{fg:struct2}
\vspace{-0.5cm}
\end{figure}

In particular, we consider a parallel algorithm that can be decomposed into multiple computations in parallel, as illustrated in Fig.~\ref{fg:struct2} (a). This parallel structure can capture many applications
where the algorithm of computation can be parallelized. For example, for graphic rendering~\cite{munshi08}, a graphic can be partitioned into multiple segments such that each segment can be processed independently. Moreover, even for algorithms that cannot be fully parallelized, they often can be partly parallelized. For instance, for image classification using a DNN model, although different layers of
the model have to be processed in order, many layers can be parallelized separately.







\noindent\textbf{Computation.} A computation is to take some data as input, execute some instructions (e.g., arithmetic operation, comparison, branching such as "if...then...") based on the input data, and produces some data as output. The \textit{workload} of a computation (e.g., the number of arithmetic operations) generally varies for different computations. Some computations cannot be executed in parallel and have to be executed in order. This is the case when the output of a computation is used as the input of another computation, such that the latter computation cannot start until the former computation is completed. The precedence relations between the computations of an algorithm can be represented by a directed acyclic graph as illustrated in Fig.~\ref{fg:struct2} (a).

We assume that the total computation workload $w$ of the algorithm is divisible, such that it can be divided into any workloads $w_i$, $\forall i\in\ca{N}$ of the parallel computation $i$, $\forall i\in\ca{N}$ with $\sum_{i\in\ca{N}}w_i=w$. For ease of exposition, without loss of generality, we assume that the workloads of computation 0 and $N+1$ are 0. The results of this paper can be used to find approximate optimal solutions when the total workload $w$ is not arbitrarily divisible (which we will study in future work).



\noindent\textbf{Communication.} For two computations that have to be executed in order, the output of one computation has to be communicated to another computation as the input. The message passing between two computations is referred to as a communication. The \textit{workload} of a communication (typically captured the amount of data to be transferred) generally varies for different communications.

The communications in the algorithm consists of \textit{forward} communications (from computation 0 to each $i\in\ca{N}$) and \textit{backward} communications (from each computation $i\in\ca{N}$ to $N + 1$). We assume that the workloads of forward and backward communications are independent of the computation workloads $w_i$, $i\in\ca{N}$\footnote{We will study the situation where the workload of a communication depends on that of the corresponding computation (e.g., proportional) in future work. We conjecture that the results for this setting will be similar to those in this paper.}. This is the case when the amounts of input and output data of a computation are independent of the computation workloads. For example, consider an algorithm that computes $\max_{\mathbf{a}\in\{0,1\}^M}f(\mathbf{a})$ by calculating $f(\mathbf{a})$ for all $\mathbf{a}\in\{0,1\}^M$ and finding the maximum (e.g., to solve a combinatorial optimization problem). The total computation workload of this algorithm is (approximately) $w=2^M$ which is the number of evaluations of $f(\mathbf{a})$. Then each computation $i$ can be designed as to compute $\max_{\mathbf{a}\in\ca{A}_i}f(\mathbf{a})$ by calculating $f(\mathbf{a})$ for all $\mathbf{a}$ in a subset $\ca{A}_i\subset\{0,1\}^M$ and finding the maximum, where $\cup_{i\in\ca{N}}\ca{A}_i=\{0,1\}^M$ and $\ca{A}_i\cap\ca{A}_j=\emptyset$, $\forall i\neq j$. Thus the workload of computation $i$ is $w_i=|\ca{A}_i|$. Therefore, for computation $i$, the input data is the specification of function $f$ and subset $\ca{A}_i$ (which can be  a set of consecutive binary sequences, specified by the smallest and largest binary sequences), and the output data is the maximum value of $f$ over $\ca{A}_i$, which are both independent of the computation workload $w_i$.



\noindent\textbf{Computing device.} We consider a set of edge devices $\ca{N}\triangleq\{1,2,\cdots,N\}$ (referred to as ``nodes'') that are available for executing the distributed algorithm. The computation rate of a device is the
computation workload that the device can complete per unit time, which quantifies the computation capability (depending on e.g., CPU, memory) of the device. It generally varies for different devices. The computations of the algorithm are allocated to edge devices as in Fig.~\ref{fg:struct2} (b).



\noindent\textbf{Edge network.} The edge devices are typically connected by wireless links. Due to interference among wireless links, only wireless links without mutual interference can transmit data concurrently.
The communication rate of a wireless link is the communication workload that the link can complete per unit time, which quantifies the communication capability (depending on e.g., transmit power, channel state) of the link. It generally varies for different wireless links.


We consider a single-hop wireless network such that each node can transmit data to each other node directly. The network is subject to complete interference constraints such that only one node can transmit at a time (i.e., no more than one node can transmit concurrently). This is a reasonable setting when nodes are close to each other, which is usually the case in an edge network (e.g., WiFi). We assume that the computation rates of devices and communication rates of wireless links are known\footnote{We will study the situation where computation and communication rates are unknown and stochastic in future work, based on the results of this paper.} (which can be estimated before executing the algorithm). As a result, the delays of executing computations and communications are known. We also assume that the communications between nodes are coordinated by a central controller (e.g., a WiFi AP), such that there is no contention or interference in the network. This can be achieved, e.g., using the point coordination function protocol of WiFi.



\noindent\textbf{Algorithm delay.} The delay for executing the distributed algorithm is the total time it takes to complete all the computations and communications of the algorithm, subject to the precedence constraints among  computations and communications.








Based on the framework described above, our goal is to solve the following problem.
\begin{df}[The problem of minimizing algorithm delay]
We aim to optimize allocating the computation workloads $w_i$, $\forall i\in\ca{N}$ of the distributed algorithm to computing nodes $i$, $\forall i\in\ca{N}$, and scheduling the communications between the nodes in the wireless network, in order to minimize the delay of executing the distributed algorithm.

\end{df}

\section{Optimal Computation Allocation and Optimal Communication Scheduling for Minimizing Algorithm Delay}\label{sc:alloc_sched}


In this section, based on the framework and problem formulation in Section~\ref{sc:mod}, we study the optimal allocation of computation workloads and the optimal scheduling of communications that minimize the delay of executing the distributed algorithm. We assume that all the available nodes are used for executing the algorithm. In Section~\ref{ssc:select}, we will relax this assumption and study the optimal selection of nodes that minimizes the algorithm delay, based on the optimal computation allocation and communication scheduling. Note that after the optimal computation allocation and communication scheduling are determined, the optimal scheduling of computations on the device nodes can be easily determined: each computation starts once its corresponding forward communication finishes.


We note that there is non-trivial interdependence between computation allocation and communication scheduling: the optimal design for one problem depends on the design for the other problem. In the following, we will first show some general structural properties that are satisfied by the optimal communication scheduling. Then, given any communication scheduling policy with these properties, we will characterize the optimal computation allocation. Next, based on the optimal computation allocation policies, we will characterize the optimal communication scheduling.

\subsection{Structural Properties of Optimal Communication Scheduling}\label{ssc:struct}


In general, a communication scheduling policy can be preemptive such that the network can interrupt the execution of a communication at any time and start to execute another communication~\cite{pinedo12}. However, we can show that it suffices to focus on non-preemptive policies.
\begin{lm}\label{lm:non-pre}
Non-preemptive communication scheduling policies are optimal.
\end{lm}


Due to space limitation, only the proofs of some results in this paper are provided (in the appendix). Lemma~\ref{lm:non-pre} shows that it is not beneficial to preempt an ongoing communication to execute another communication. Intuitively, this is because preemptive scheduling is typically better than non-preemptive scheduling when tasks become available at different times and the objective is to minimize the total delay of tasks~\cite{pinedo12}. In contrast, for the problem here, the communications are always available (subject to that a backward communication is after the corresponding forward communication), and the objective is to minimize the algorithm delay which is equal to the maximum delay of communication.


Then we show that it is optimal to schedule all the forward communications before all the backward communications.
\begin{lm}\label{lm:order}
It is optimal to schedule all the forward communications before all the backward communications.
\end{lm}

Lemma~\ref{lm:order} provides the insight that it is always beneficial to schedule any forward communication before any backward communication compared to the other way, as it allows for more time to execute the computations associated with these communications.


\begin{figure}
\centering
\includegraphics[width=0.33\textwidth]{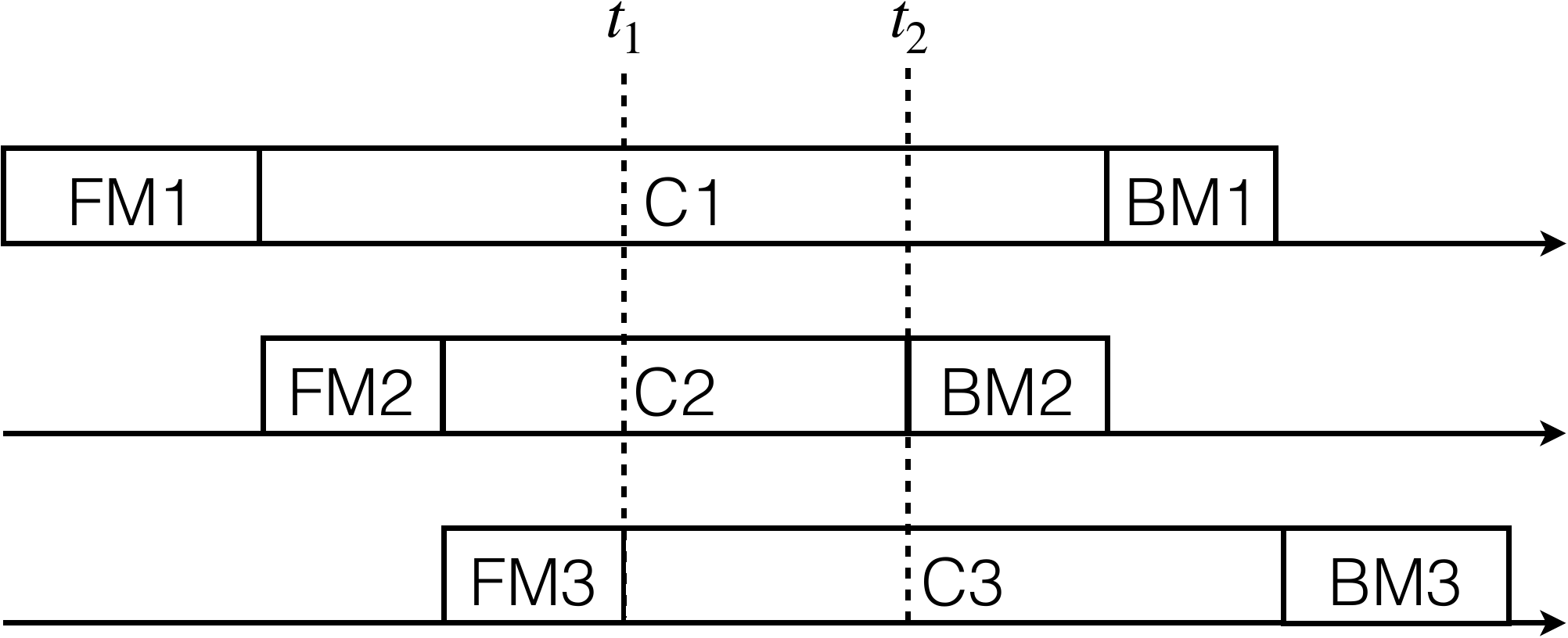}
\caption{A Computation is represented by ``C'' and a coMmunication is represented by ``M''. A forward communication is represented by ``F'' and a backward communication is represented by ``B''. Each axis shows the schedule of computations and communications for a node.}
\label{fg:alloc}
\vspace{-0.5cm}
\end{figure}

Next we show that it is optimal for the wireless network to keep busy between forward communications and between backward communications.
\begin{lm}\label{lm:non-idle}
It is optimal for the communication scheduling policies to be non-idle between forward communications and between backward communications, respectively.
\end{lm}

The non-idle optimal policy in Lemma~\ref{lm:non-idle} means that the wireless network has no idle period between any two forward communications and between any two backward communications. However, there can be some idle period between the last forward communication and the first backward communication (i.e., between time $t_1$ and time $t_2$ in Fig.~\ref{fg:alloc}). Lemma~\ref{lm:non-idle} provides the insight that the wireless network should keep performing communications without any idle period, so as to complete communications as soon as possible, unless it is necessary to wait for some period during which the nodes can perform computations. 


\subsection{Optimal Computation Allocation}\label{ssc:alloc}

In this subsection, we study the optimal allocation of computation workloads to nodes given any communication scheduling policy that satisfies the structural properties discussed in Section~\ref{ssc:struct}. 


The optimal computation allocation can be found by an efficient algorithm that consists of up to three phases as described in Algorithm~\ref{al:alloc}. In particular, in Phase 1, we first allocate computation workloads as much as possible to nodes such that all these workloads can be completed before the last forward communication ends (i.e., before time $t_1$ in Fig.~\ref{fg:alloc}). If the total workload of the algorithm can be fully allocated in this way, we have found the optimal computation allocation. Otherwise, in Phase 2, we allocate computation workloads as much as possible to nodes such that all these workloads can be completed after the first backward communication starts (i.e., after time $t_2$ in Fig.~\ref{fg:alloc}). If the remaining workload of the algorithm can be fully allocated in this way, we have found the optimal computation allocation. Otherwise, in Phase 3, we allocate the further remaining workload of the algorithm to nodes such that it can be completed after the last forward communication ends and before the first backward communication starts (i.e., between time $t_1$ and time $t_2$ in Fig.~\ref{fg:alloc}). It can be seen that the computational complexity of Algorithm~\ref{al:alloc} is $O(N)$, as each phase of the algorithm involves at most $N$ iterations. We establish the optimality of Algorithm~\ref{al:alloc} as follows.
\begin{ps}\label{ps:alloc}
For any communication scheduling policy that satisfies the structural properties in Lemmas~\ref{lm:non-pre}, \ref{lm:order}, and~\ref{lm:non-idle}, Algorithm~\ref{al:alloc} finds the optimal allocation of computation workloads to nodes.
\end{ps}



Proposition~\ref{ps:alloc} provides some interesting insights regarding the optimal computation allocation characterized by Algorithm~\ref{al:alloc}. Intuitively, the optimal policy should reduce the idle computing periods of nodes as much as possible, so as to minimize the algorithm delay. To this end, it allocates workloads to the idle period of each node after its forward communication ends and before the last forward communication (among all nodes) ends, and to the idle period after the first backward communication (among all nodes) starts and before its backward communication starts, until there is no such idle period. The workloads allocated to these idle periods do not increase the algorithm delay, as it remains equal to the total delay of all forward and backward communications. If there is some workload of the algorithm that remains unallocated after the above allocation, it is allocated to all nodes in proportional to their computation rates, such that it incurs an equal computation delay to all the nodes. This delay increases the algorithm delay beyond the delay incurred by communications. As a result of Proposition~\ref{ps:alloc} and Algorithm~\ref{al:alloc}, we can see that when the total workload of the algorithm is sufficient (above some threshold), each node keeps performing its computation between its forward and backward communications. Otherwise, some node is forced to be idle between its forward and backward communications.


\begin{algorithm}[t]
\textbf{input:} orders of forward and backward communications $(I_1,\cdots,I_N)$ and $(J_1,\cdots,J_N)$, delays of forward and backward communications $s_i$ and $d_i$, $\forall i\in\ca{N}$, total computation workload $w$\\
$w^*_i=0$, $\forall i\in\ca{N}$, $j=1$\;
\texttt{// Phase 1}\;
\While {$w>0$ and $j\le N-1$}{
$w^1_j=\min(r_j\sum^{I_N}_{i=I_{j+1}}s_i,w)$\;
$w=w-w^1_j$\;
$j=j+1$\;}
\texttt{// Phase 2}\;
$j=N-1$\;
\While {$w>0$ and $j\ge 2$}{
$w^2_j=\min(r_j\sum^{J_j}_{i=J_1}d_i,w)$\;
$w=w-w^2_j$\;
$j=j-1$\;}
\texttt{// Phase 3}\;
\If {$w>0$}
{\ForEach {$i\in\ca{N}$}{$w^3_i=wr_i/\sum_{i\in\ca{N}}r_i$\;}}
\ForEach {$i\in\ca{N}$}{$w^*_i=w^1_i+w^2_i+w^3_i$\;}
\textbf{output:} optimal computation workload $w^*_i$ of each node $i$, $\forall i\in\ca{N}$
\caption{Find the optimal computation allocation}\label{al:alloc}
\end{algorithm}


\subsection{Optimal Scheduling Order of Communications}\label{ssc:order}


In this subsection, based on the structural properties of the optimal communication scheduling in Section~\ref{ssc:struct} and the optimal computation allocation in ~\ref{ssc:alloc}, we study the optimal scheduling order of communication. Due to the symmetry between forward communications and backward communications, we focus on the scheduling order of backward communications, as the results for forward communications follow similarly. In particular, we consider the optimal scheduling order that minimizes the algorithm delay, given the total computation workload $w$ of the algorithm. Based on the optimal computation allocation found by Algorithm~\ref{al:alloc}, we can transform this problem to an equivalent ``dual'' form: how to schedule the backward communications, such that the total computation workload $v$ that \textit{can be completed after the first backward communication starts} (i.e., after time $t_2$ in Fig.~\ref{fg:alloc}) is maximized?

\begin{figure}
\centering
\includegraphics[width=0.42\textwidth]{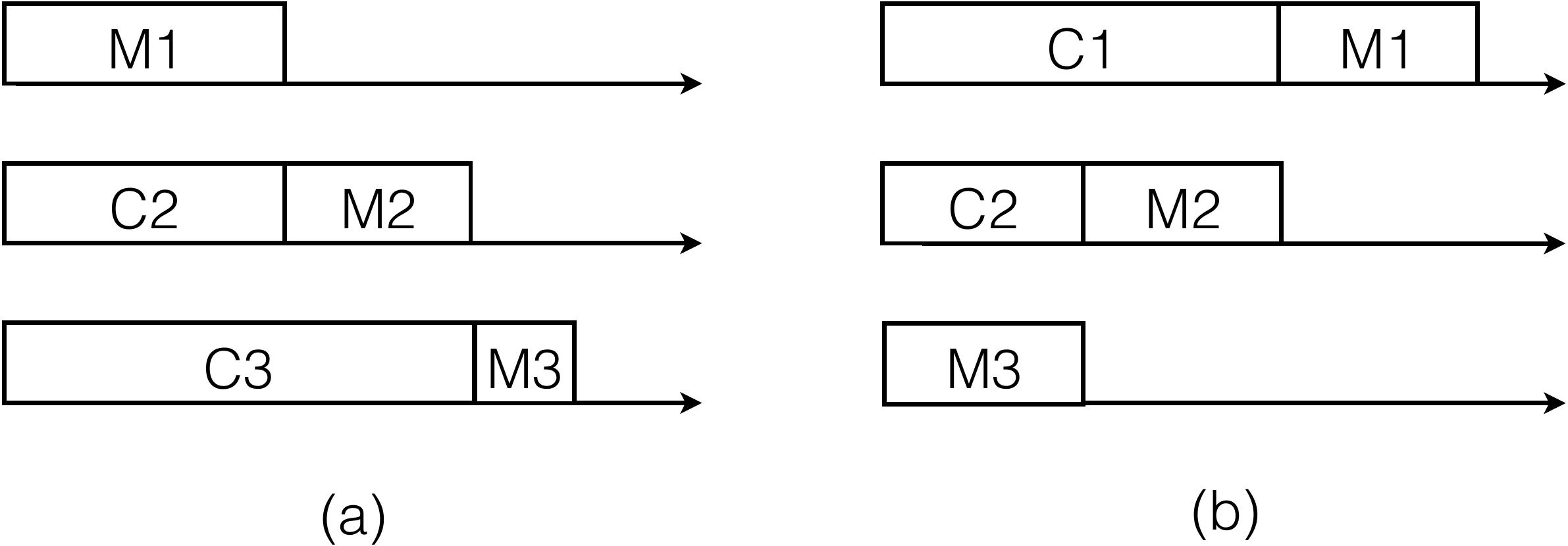}
\caption{(a) It is optimal that the longest communication M1 is scheduled first; (b) It is optimal that the fastest computing node (which is node 1) finishes its computation last.}
\label{fg:sched2}
\end{figure}


We first consider the case where nodes have uniform computation rates but can have diverse communication delays.  The optimal scheduling order is given as follows.
\begin{ps}\label{ps:div_net}
For the case of uniform computation rates, it is optimal to schedule communications in the descending order of communication delays.
\end{ps}

Proposition~\ref{ps:div_net} means that the optimal policy schedules the longest communication (i.e., with the largest delay) first, and the second longest next, etc, as illustrated in Fig.~\ref{fg:sched2}(a). This result provides the following insight: it is better to schedule a longer communication earlier than a shorter one, since it allows for more time for other node(s) to perform computing. As a result, the computing capacities of nodes are most efficiently utilized and thus the algorithm delay is minimized. 




Then we consider the case where nodes have uniform communication delays but can have diverse computation rates. The optimal scheduling order is given below.
\begin{ps}\label{ps:div_comp}
For the case of uniform communication delays, it is optimal to schedule communications in the ascending order of the corresponding nodes' computation rates.
\end{ps}

Proposition~\ref{ps:div_comp} means that the optimal policy schedules the communication of the slowest node first, and that of the second slowest node next, etc, as illustrated in Fig.~\ref{fg:sched2}(b). The insight from this result is as follows: it is better to utilize a faster node for a longer period than a slower node, so that the computing capacities of nodes are most efficiently utilized and thus the algorithm delay is minimized. 


\begin{algorithm}[t]
\textbf{input:} computation rates $r_i$, $\forall i\in\ca{N}$, communication delays $d_i$, $\forall i\in\ca{N}$\\
$v^*=0$, $I^*_i=i$, $\forall i\in\ca{N}$\;
\ForEach {$(I_1,\cdots,I_k)$ with $I_i\in\ca{N}$, $\forall i\in\{1,\cdots,k\}$ and $I_i\neq I_j$, $\forall i\neq j$}
{compute 
\[v=d_{I_1}\!\!\!\!\sum_{i\in\ca{N}\setminus\{I_1\}}\!\!\!\!r_i +d_{I_2}\!\!\!\!\sum_{i\in\ca{N}\setminus\{I_1,I_2\}}\!\!\!\!r_i+\cdots+d_{I_k}\!\!\!\!\!\!\!\!\sum_{i\in\ca{N}\setminus\{I_1,\cdots,I_k\}}\!\!\!\!\!\!\!\!r_i ;\]\
\If{$v>v^*$}{$v^*=v$, $I^*_i=I_i$, $\forall i\in\{1,\cdots,k\}$\;}
}
set $I^*_i\in\ca{N}$, $\forall i\in\{k+1,\cdots,N\}$ such that $I^*_i\neq I^*_j$, $\forall i\neq j$ and $j\in\ca{N}$\;
 
 \textbf{output:} scheduling order $(I^*_1,\cdots,I^*_N)$
 
\caption{Find an approximately optimal scheduling order of communications}\label{al:order}
\end{algorithm}


Next we consider the general case where nodes can have diverse computation rates and also diverse communication delays. It is plausible that the optimal policies in the previous two cases can perform well for the case here. However, we can find some counterexamples which show that those policies can result in a solution that is arbitrarily worse than the optimal solution (given in the appendix). 

To determine the optimal scheduling order, we can use an exhaustive search by calculating the total computation workload $v$ (that can be completed after the first backward communication starts) for all possible scheduling orders and then finding the optimal one. However, the computational complexity of the exhaustive search is $O(N!)$ which is  too high. Therefore, we aim to find a computationally efficient approximation algorithm that can provide performance guarantee in terms of the approximation ratio $v/v^*$, where $v^*$ is the total computation workload for the optimal scheduling order. It can be shown that the largest delay first and fastest node last policies (which are the optimal policies for the previous two cases, respectively) cannot provide a finite approximation ratio due to ignoring computation rates or communication delays, respectively. Thus motivated, the design of the approximation algorithm here should take into account both factors. In particular, we design a greedy algorithm that calculates the total computation workload $v$ for all possible scheduling orders of the \textit{first $k$ communications} among all the $N$ communications, where $1\le k\le N$, and then finds the optimal order of the first $k$ communications. The scheduling order of the remaining $N-k$ communications can be set arbitrarily. The algorithm is described in detail in Algorithm~\ref{al:order}. We can show that this algorithm has a finite approximation ratio as follows.

\begin{ps}\label{ps:approx}
Algorithm~\ref{al:order} finds a communication scheduling order that has an approximation ratio of $k/N$ compared to the optimal policy, i.e., $v/v^*\ge k/N$.
\end{ps}

Proposition~\ref{ps:approx} shows that there exists a tradeoff between the computational complexity of Algorithm~\ref{al:order} and its approximation ratio, which can be controlled by the parameter $k$. The complexity of the algorithm is $O({N \choose k})$ which is in the order of $O(N^k)$. Therefore, a larger $k$ means higher complexity which is worse, but also a higher approximation ratio which is better.


%

\begin{figure}
\centering
\includegraphics[width=0.45\textwidth]{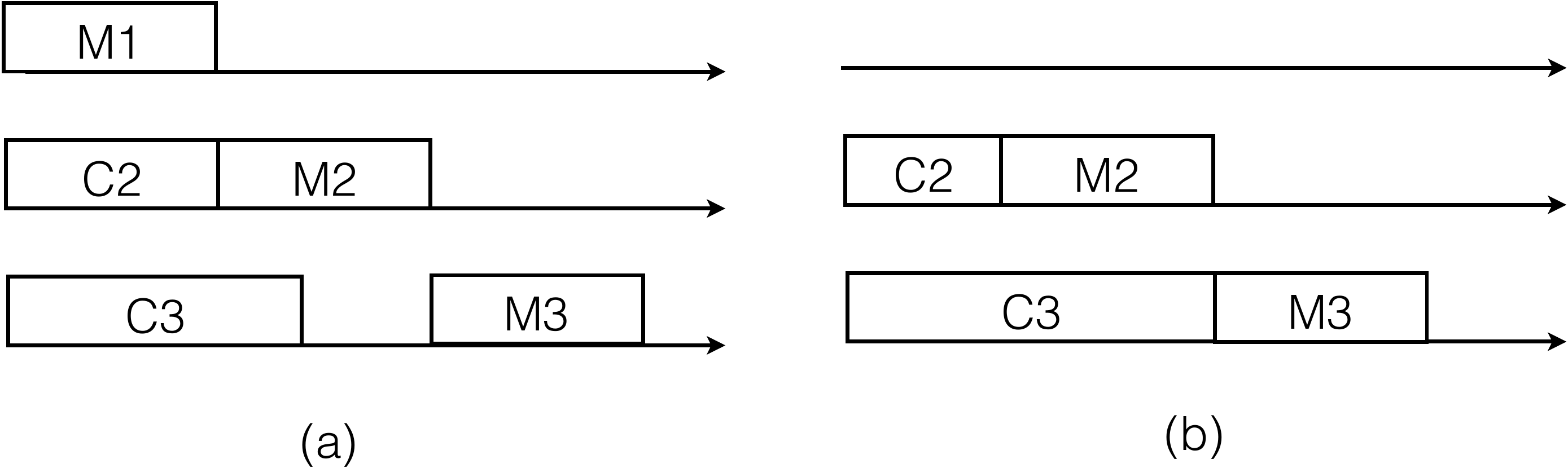}
\caption{It is optimal to use 2 nodes as in (b) rather than 3 nodes as in (a).}
\label{fg:sched3}
\vspace{-0.5cm}
\end{figure}

\subsection{Optimal Selection of Nodes}\label{ssc:select}



In the previous subsections, it is assumed that each node is used to execute the distributed algorithm, such that the forward and backward communications of that node is scheduled regardless of the computation workload allocated to it (even when no workload is allocated). However, it is important and interesting to note that it may not be optimal to use as many nodes as possible to execute the algorithm. This is because using an additional node incurs communication delays which can increase the algorithm delay. This increase can outweigh the decrease of the computation delay due to utilizing the additional node for computing, as illustrated in Fig.~\ref{fg:sched3}. Thus motivated, in the following we investigate how to select nodes for executing the algorithm to minimize the algorithm delay.


\begin{figure*}[t!]
\begin{minipage}[t]{.29\textwidth}
\includegraphics[width=1.\textwidth]{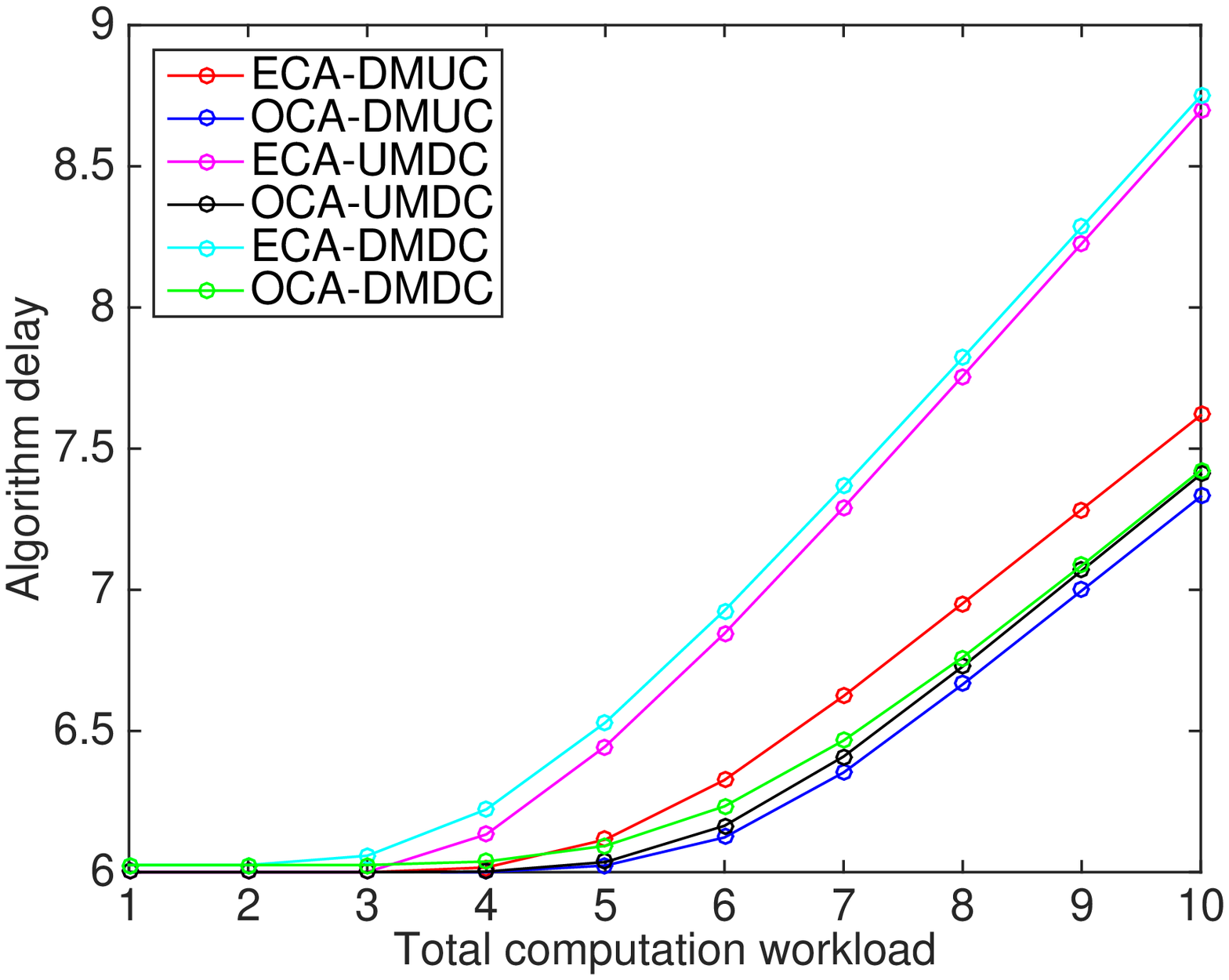}
\caption{Impact of computation allocation}
\label{fg:alloc_v}
\end{minipage}
\hfill
\begin{minipage}[t]{.29\textwidth}
\includegraphics[width=1.\textwidth]{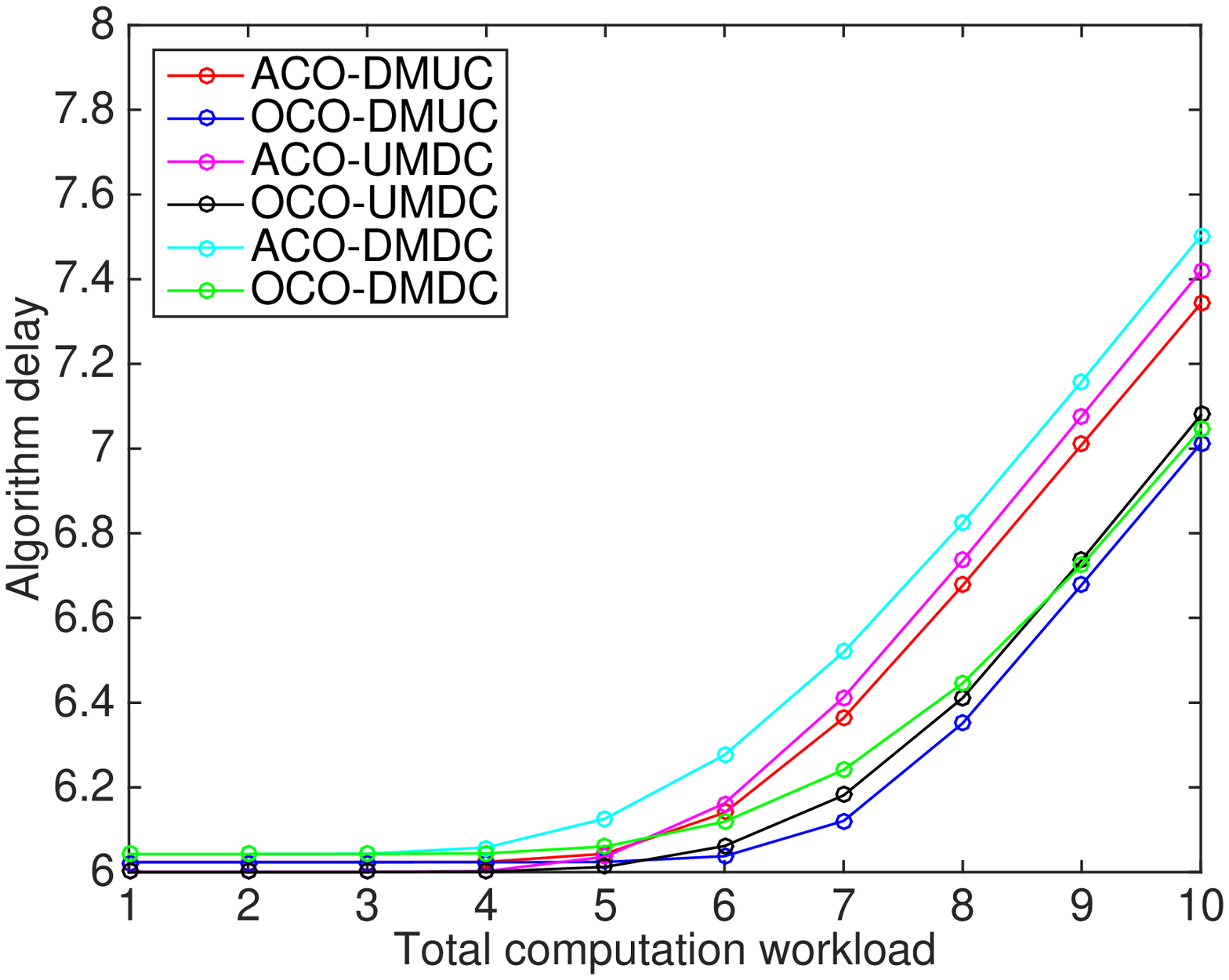}
\caption{Impact of communication order}
\label{fg:order_v}
\end{minipage}
\hfill
\begin{minipage}[t]{.29\textwidth}
\includegraphics[width=1.\textwidth]{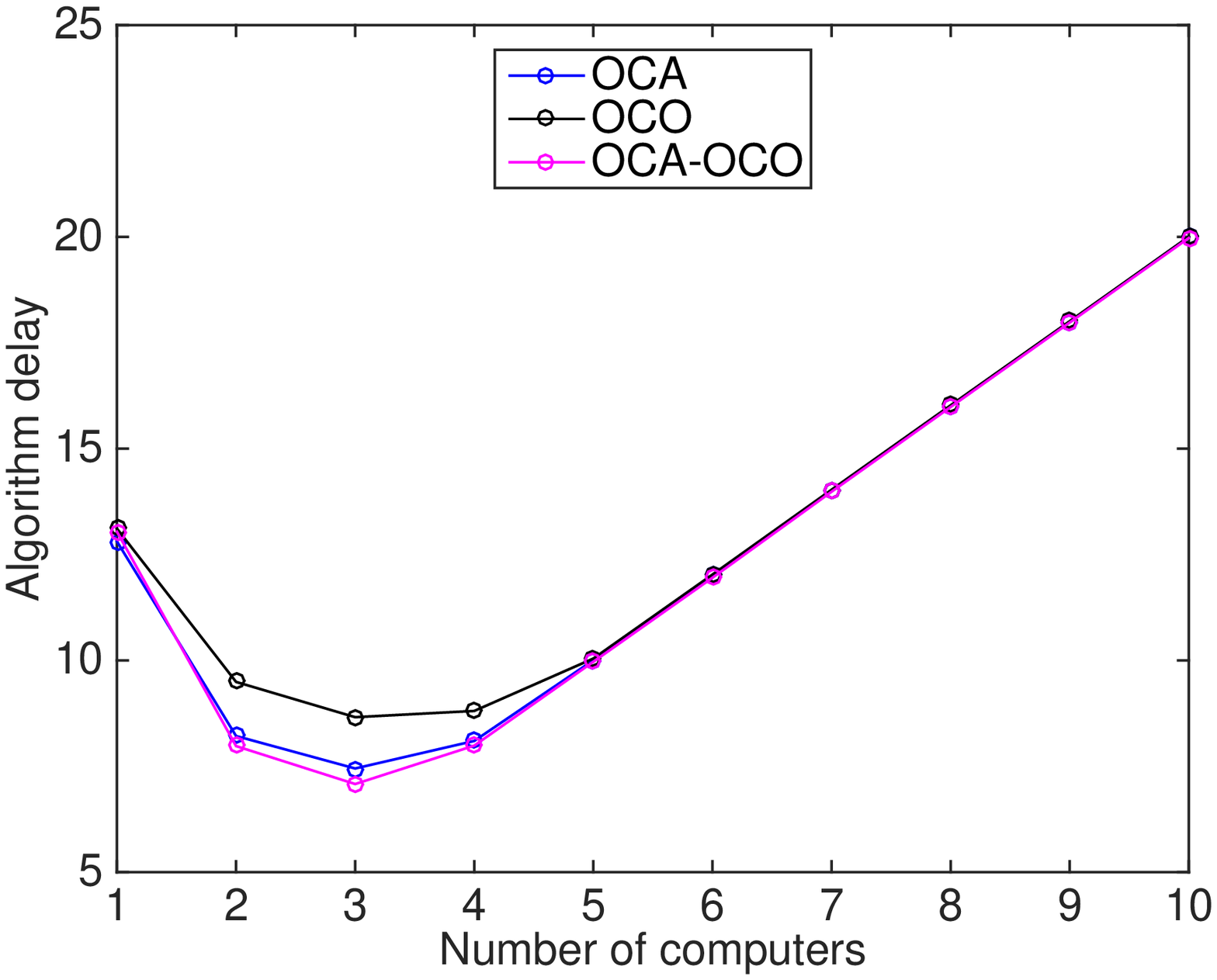}
\caption{Impact of node selection}
\label{fg:sel_N}
\end{minipage}
\vspace{-0.cm}
\end{figure*}

%
%


We first consider the cases where nodes have uniform computation rates or uniform communications delays. For these two cases, we can show that the optimal set of nodes to select is those with the smallest communication delays or the highest computation rates.
\begin{lm}\label{lm:sel}
For the cases of uniform computation rates (or uniform communications delays), if a node is in the optimal set of nodes, then each node with a smaller communication delay (or higher computation rate, respectively) is also in the optimal set.
\end{lm}

Lemma~\ref{lm:sel} is intuitive as a node with a smaller communication delay or higher computation rate should be preferred over one with a larger communication delay or lower computation rate, respectively. Based on this result, we can use an efficient linear exhaustive search to find the optimal set of nodes. In particular, it calculates the minimum delay for all possible numbers of the ``best'' nodes (i.e., the $k$ nodes with the smallest communication delays or highest computation rates where $k\in\{1,2,\cdots,N\}$), and then finds the optimal number of the ``best'' nodes. The computational complexity of the linear exhaustive search is $O(N)$.



Then we consider the general case where nodes can have diverse computation rates and also diverse communication delays. To determine the optimal set of nodes, we may use an exhaustive search that calculates the (approximate) minimum delay for all possible sets of nodes (as the optimal scheduling order of communications is difficult to find), and then finds the optimal set among them. However, the computational complexity of the exhaustive search is $O(2^N)$ which is too high. Therefore, we can use a greedy algorithm instead as follows: we start with the empty set, and in each iteration we add to this set the node not selected that can reduce the algorithm delay the most, until no such node exists. The computational complexity of this greedy algorithm is $O(N)$. We will analyze the performance of this algorithm in terms of its approximation ratio in our future work.

%
%
%

\section{Simulation Results}\label{sc:sim}

In this section, we evaluate the performance of the optimal computation allocation, the optimal communication order, and the optimal selection of nodes using simulation results.

\subsection{Optimal Computation Allocation}



To illustrate the efficiency of the optimal computation allocation, we compare the algorithm delay under the optimal computation allocation (OCA) found by Algorithm~\ref{al:alloc}, and under equal computation allocation (ECA) that allocates an equal computation workload $w/N$ to each node. We set the default parameters as follows: $N=3$, $w=10$, $s_i=1$, $d_i=1$, $r_i=1$, $\forall i$.

Fig.~\ref{fg:alloc_v} illustrates the algorithm delay under OCA and under ECA, when the total computation workload $w$ of the algorithm varies. We can see that the delay under OCA is always no greater than that under ECA, which demonstrates the better performance of OCA. We can also see that the delay is non-decreasing with $w$, which is because a larger workload takes more time to complete. We note that the performance gap is 0 when $w$ is small. This is because in this case, all the workloads can be completed before the last forward communication ends or after the first backward communication starts, such that the delay is equal to the total time of forward and backward communications, which is the same for both allocations. We further observe that the performance gain of OCA compared to ECA for diverse communication delays and computation rates (DMDC) is more than that for uniform communication delays and diverse computation rates (UMDC), or for diverse communication delays and uniform computation rates (DMUC). This is because when communication delays or computation rates are diverse rather than uniform, OCA is more different from ECA, so that OCA is more beneficial.



\subsection{Optimal Communication Order}


To illustrate the efficiency of the optimal communication scheduling order, we compare the algorithm delay under the optimal order given by Propositions~\ref{ps:div_net} or~\ref{ps:div_comp}, or the approximate optimal order found by Algorithm~\ref{al:order}, and under an arbitrary order that schedules communications in the ascending order of users' indices.

Fig.~\ref{fg:order_v} illustrates the algorithm delay under the (approximate) optimal communication order (OCO) and under the arbitrary communication order (ACO), as the the total computation workload $w$ of the algorithm varies. As expected, we can see that OCO always outperforms ACO, and the delay is non-decreasing with $w$. Similar to Fig.~\ref{fg:alloc_v}, we note that the performance gap is 0 when $w$ is small, which is because in this case the delay is equal to the total delay of communications, which is the same for both scheduling orders. We can also observe that the performance gain of OCO compared to ACO for DMDC is more than that for UMDC or DMUC. Similar to Fig.~\ref{fg:alloc_v}, the reason is that when communication delays or computation rates are diverse rather than uniform, OCO is more different from ECA and thus is more beneficial.


\subsection{Optimal Selection of Nodes}


To illustrate the impact of the selection of nodes, we compare the algorithm delay as the number of nodes selected for executing the algorithm varies. Fig.~\ref{fg:sel_N} illustrates the algorithm delay under the optimal computation allocation (OCA), under the (approximate) optimal communication order (OCO), and under both (OCA-OCO), as the number of selected nodes $N$ varies. We can see that the delay first decreases and then increases with $N$. This is because the delay reduction due to the computation workload completed by an additional node first outweighs the delay increase due to the communications of that node, and then the former effect is dominated by the second effect. We can also observe that the delay under OCA-OCO is better than under OCA or OCO only, which demonstrates that both OCA and OCO are beneficial.

\section{Conclusion and Future Work}\label{sc:con}


In this paper, we have explored DEC by studying the minimization of the delay of executing a distributed algorithm using distributed edge devices connected by a wireless network. In particular, we have characterized the optimal communication scheduling, the optimal computation allocation, and the optimal selection of nodes for minimizing the algorithm delay. The optimal policies have been developed by addressing the non-trivial coupling between these three issues, while taking into account the features of wireless networks. The results have provided useful insights into the optimal policies.


For future work, one immediate direction is to investigate the case where communication and computation delays are unknown and stochastic. Another important setting is when the computation workload of the algorithm is not arbitrarily divisible. We will also study the setting when the workloads of communications depend the corresponding workloads of computations. We will further explore distributed algorithms with other structures, including a serial structure, and a combination of serial and parallel structures.





\section*{APPENDIX}

\subsection*{Counterexamples of communication scheduling orders}

Next we present counterexamples where the communication scheduling orders given by the largest delay first (LDF) policy and the fastest computing node last (FCL) policy are arbitrarily worse than the optimal orders, for the case of diverse computation rates and diverse communication delays.


First consider a setting where the delays of the communications are in the order $t_1\ge t_2\ge\cdots \ge t_{N-1}\ge t_N$, such that the LDF policy results in the order $(1, 2,\cdots,N-1,N)$. Then the maximum computation workload that can be completed under this scheduling order is given by
\begin{align}\label{naive_load}
w=t_1\sum^N_{i=2}r_i+\cdots+s_j\!\!\sum^N_{i=j+1}r_i+\cdots+t_{N-1}r_N.
\end{align}
However, the optimal scheduling order can be in the form of $(2,3,\cdots,N-1,N, 1)$, such that the maximum workload is given by
\begin{align}\label{opt_load}
w^*=t_2\!\left(\!\!r_1\!\!+\!\!\sum^{N-1}_{i=3}\!\!r_i\!\!\right)\!\!+\!\cdots\!+\!s_j\left(\!\!r_1\!\!+\!\!\!\sum^{N-1}_{i=j+1}\!\!\!r_i\!\!\right)\!+\!\cdots+\!\!t_{N}r_1,
\end{align}
if $r_1$ is sufficiently large. Furthermore, we can see that if $r_1\rightarrow\infty$, we have $w/w^*\rightarrow0$, which means that the solution $w$ given by the LDF policy can be arbitrarily worse than the optimal solution $w^*$. This is because LDF ignores the computation rates $\{r_i\}$ in determining the scheduling order, so that the maximum workload $w$ is independent of the computation rate $r_1$, which can be very large. Consider another example where the nodes' computation rates are in the order $r_1\le r_2\le\cdots \le r_{N-1}\le r_N$, such that the FCL policy results in the order $(1, 2,\cdots,N-1,N)$. Then the maximum computation workload that can be completed is also given by~\eqref{naive_load}. However, the optimal scheduling order can also be $(2,3,\cdots,N-1,N, 1)$, such that the maximum workload is also given by~\eqref{opt_load}, if $t_N$ is sufficiently large. Then we can see that if $t_N\rightarrow\infty$, we have $w/w^*\rightarrow0$, which means that the solution $w$ given by FCL can be arbitrarily worse than the optimal solution $w^*$. This is because FCL ignores the communication delays $\{s_i\}$ in determining the scheduling order, so that the maximum workload $w$ is independent of the communication delay $t_N$, which can be very large.


\subsection*{Proof of Lemma~\ref{lm:non-pre}}

\begin{figure}
\centering
\includegraphics[width=0.42\textwidth]{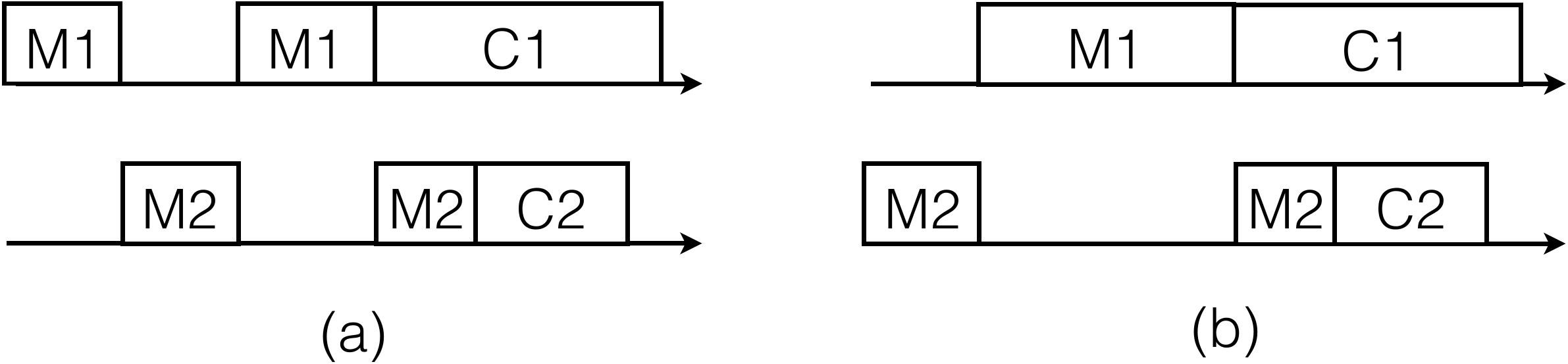}
\caption{An exchange argument for the proof of Lemma~\ref{lm:non-pre}.}
\label{fg:nonpre_pf}
\vspace{-0.3cm}
\end{figure}

The main idea of the proof is an exchange argument. WLOG, suppose forward communication M1 is interrupted by forward communication M2 into two parts, such that the 2nd part of M1 starts after M2 (or part of M2) is completed, as illustrated in Fig.~\ref{fg:nonpre_pf} (a). If M1 is interrupted for multiple times, the proof follows by applying the exchange argument multiple times. Now we exchange the scheduling order of the 1st part of M1 and M2 (or the interrupting part of M2), as illustrated in Fig.~\ref{fg:nonpre_pf} (b). We can see that the order exchange does not affect the schedules of computations C1 and C2, as well as the schedule of any communication or computation on the nodes other than nodes 1 and 2. As a result, the delay of the algorithm remain the same. This completes the proof. 

\subsection*{Proof of Lemma~\ref{lm:order}}

\begin{figure}[t]
\centering
\includegraphics[width=0.42\textwidth]{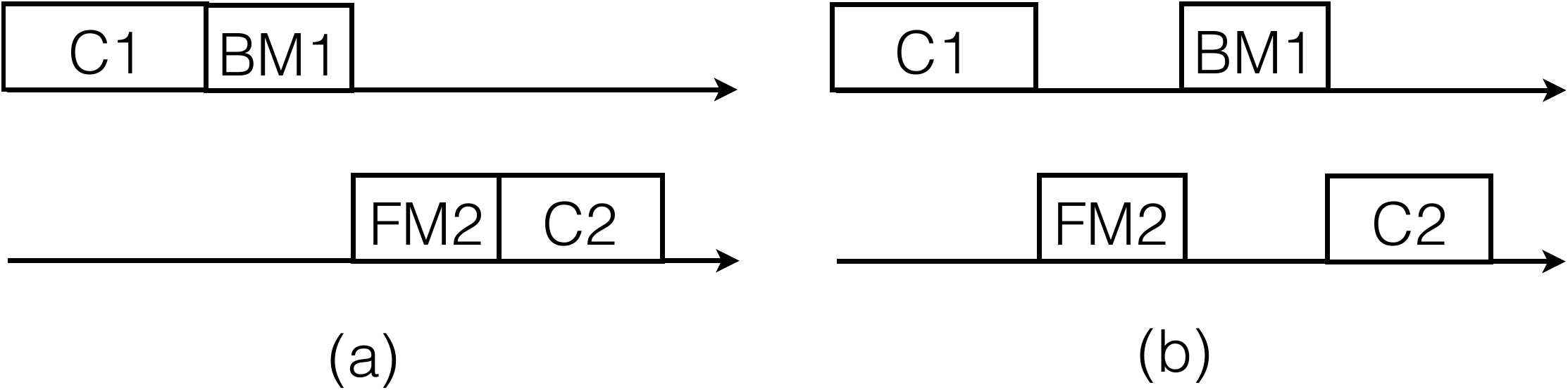}
\caption{An exchange argument for the proof of Lemma~\ref{lm:order}.}
\label{fg:order_pf}
\end{figure}

The main idea of the proof is an exchange argument. According to Lemma~\ref{lm:non-pre}, it suffices to focus on non-preemptive scheduling. WLOG, suppose forward communication FM1 is scheduled after backward communication BM2, as illustrated in Fig.~\ref{fg:order_pf} (a). Now we exchange the scheduling order of FM1 and BM2, as illustrated in Fig.~\ref{fg:order_pf} (b). We can see that the order exchange does not affect the schedules of computations C1 and C2, as well as the schedule of any communication or computation on the nodes other than nodes 1 and 2. As a result, the delay of the algorithm remain the same. This completes the proof. 

\subsection*{Proof of Lemma~\ref{lm:non-idle}}

\begin{figure}[t]
\centering
\includegraphics[width=0.42\textwidth]{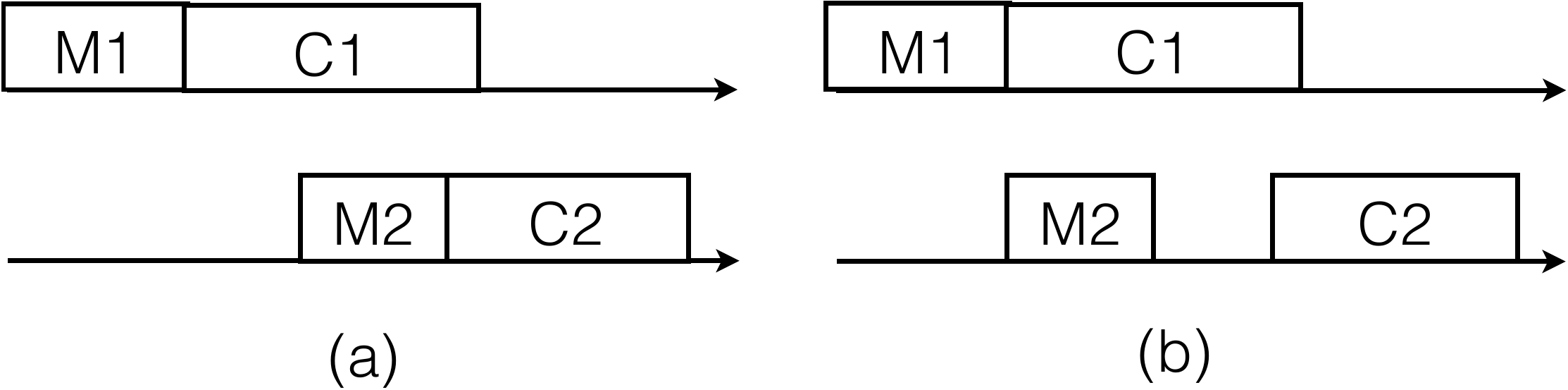}
\caption{A shifting argument for the proof of Lemma~\ref{lm:non-idle}.}
\label{fg:nonidle_pf}
\vspace{-0.3cm}
\end{figure}

The main idea of the proof is a shifting argument. According to Lemma~\ref{lm:non-pre} and Lemma~\ref{lm:order}, it suffices to focus on non-preemptive scheduling policies that schedule all forward communications before all backward communications. WLOG, suppose there is an idle period of the wireless network between forward communication M1 and forward communication M2, as illustrated in Fig.~\ref{fg:nonidle_pf} (a). If there are multiple idle periods, the proof follows by applying the exchange argument multiple times. Now we shift M2 to be right after M1, as illustrated in Fig.~\ref{fg:nonidle_pf} (b). We can see that the shifting does not affect the schedules of computations C1 and C2, as well as the schedule of any communication or computation on the nodes other than nodes 1 and 2. As a result, the delay of the algorithm remain the same. If M1 and M2 are backward communications, we can shift M1 to be right before M2, and the same argument applies. This completes the proof.

\subsection*{Proof of Proposition~\ref{ps:alloc}}

\begin{figure}[t]
\centering
\includegraphics[width=0.48\textwidth]{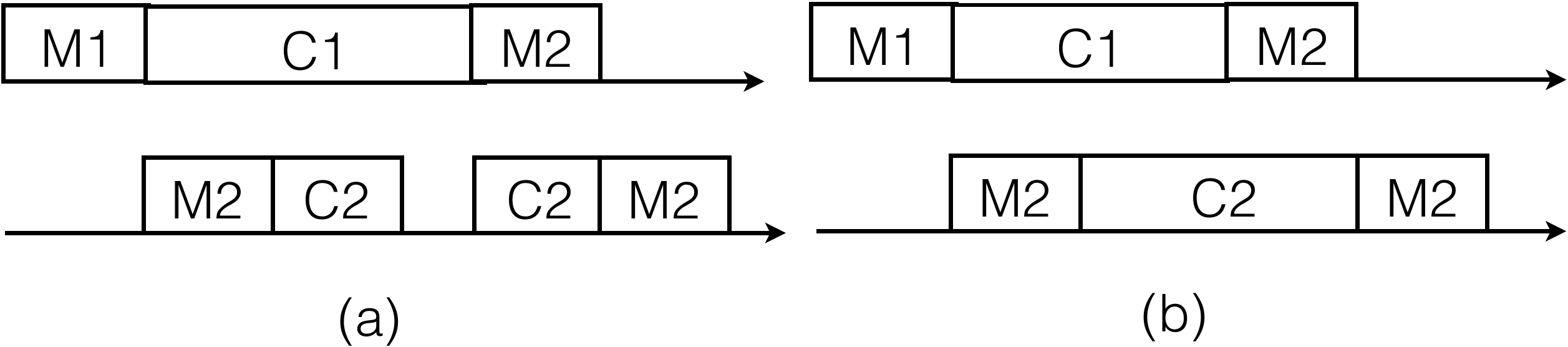}
\caption{A shifting argument for the proof of Proposition~\ref{ps:alloc}.}
\label{fg:alloc_pf}
\vspace{-0.3cm}
\end{figure}

The main idea of the proof is a shifting argument. We first note that a lower bound of the algorithm delay is the total delay $D$ of all forward and backward communications. Therefore, if Algorithm~\ref{al:alloc} terminates in Phase 1 or Phase 2, then it finds a feasible schedule of all the computation workloads of the algorithm, such that the algorithm delay is equal to $D$. Suppose the Algorithm~\ref{al:alloc} terminates in Phase 3, and the remaining total unallocated computation workloads is not allocated to the nodes in proportional to their computation rates. Then there must be some node that is idle for some period after all forward communications and before all backward communications, as illustrated in Fig.~\ref{fg:alloc_pf} (a). In this case, we can always shift some workload from some other node to this node without increasing the algorithm delay, until there is no such idle period, as illustrated in Fig.~\ref{fg:alloc_pf} (b). This completes the proof.

\subsection*{Proof of Proposition~\ref{ps:approx}}

Let $o$ be the scheduling order found by Algorithm~\ref{al:order} such that the communication of node $o(i)$ is scheduled at the $i$th in the order. Let $o^*$ be the optimal scheduling order. For convenience, define 
\[f_i\triangleq t_{o(i)}\sum^N_{j=i+1}r_{o(j)}, \ \ f^*_i\triangleq t_{o^*(i)}\sum^N_{j=i+1}r_{o^*(j)}.\]
Also define $v\triangleq\sum^N_{i=1}f_i$ and $v^*\triangleq\sum^N_{i=1}f^*_i$. Let $o_k$ be an ascending order of $k$ elements in $\mathcal{N}$. For any $o^*$ and $o_k$, we construct a new order $o'$ of the elements in $\mathcal{N}$ such that $o'(i)=o^*(o_k(i))$, $\forall i\in\{1,\cdots,k\}$ (the elements $o'(i)$, $\forall i\in\{k+1,\cdots,N\}$ can be arbitrary, so there can be multiple such $o'$). Intuitively, we move the $k$ elements of $o^*$ with order indices given by $o_k$ to be the first $k$ elements in $o'$, while keeping the relative order of these $k$ elements the same. For example, if $o^*=(5,4,3,2,1)$ and $o_k=(1,3,5)$ with $k=3$, then $o'=(5,3,1,4,2)$. Define
\[f'_i\triangleq t_{o'(i)}\sum^N_{j=i+1}r_{o'(j)}.\]
Then we have
\begin{align}\label{approx_pf}
f'_{i}=t_{o'(i)}\sum^N_{j=i+1}r_{o'(j)}&=t_{o^*(o_k(i))}\sum^N_{j=i+1}r_{o'(j)}\nonumber\\
&\ge t_{o^*(o_k(i))}\!\!\!\!\sum^N_{j=o_k(i)+1}\!\!\!\! r_{o^*(j)}=f^*_{o_k(i)} 
\end{align}
where the inequality is due to that 
\[\{o^*(o_k(i)+1), o^*(o_k(i)+2), \cdots, o^*(N)\}\]
is a subset of 
\[\{o'(i+1), o'(i+2), \cdots, o'(N)\}.\]
To see this, for the previous example, if $i=2$, we have
\[\{o^*(4),o^*(5)\}=\{2,1\}\subset \{o'(3), o'(4), o'(5)\}=\{1,4,2\}.\]

Then we have
\begin{align*}
\sum^k_{i=1}f_i\ge \sum^k_{i=1}f'_{i}\ge \sum^k_{i=1}f^*_{o_k(i)}
\end{align*}
for any $o_k$, where the 1st inequality is due to Algorithm~\ref{al:order}, and the 2nd inequality is due to~\eqref{approx_pf}. Using the above we have
\begin{align*}
{N\choose k}v= {N\choose k} \sum^k_{i=1}f_i &\ge {N\choose k}  \sum^k_{i=1}f^*_{o_k(i)}\\
&= {N\choose k}\frac{k}{N}\sum^N_{i=1}f^*_i={N\choose k}kv^*/N
\end{align*}
where the 2nd equality is due to that we sum up $\sum^k_{i=1}f^*_{o_k(i)}$ for all possible orders of $o_k$ whose number is ${N\choose k}$. Thus we have $v/v^*\ge k/N$. This completes the proof. 

\newpage
\bibliographystyle{IEEEtran}
\bibliography{IEEEabrv,edge,distributed,network,book,newbib}


\end{document}